\documentclass[9pt]{book}

\usepackage[dvips]{graphicx,color}
\usepackage{makeidx,universe}

\makeauthorindex

\BookTitle{The proceedings of the Physics of Accreting Compact Binaries}

\CopyRight{\copyright 2010 by Universal Academy Press, Inc.}

\begin{document} 

\pagenumbering{arabic}

\chapter{%
FS Aurigae: a Triple Cataclysmic Variable System \\containing Precessing, Magnetic White Dwarf}

\author{\raggedright \baselineskip=10pt%
{\bf G. Tovmassian,$^{1}$
S. Zharikov,$^{1}$
C. Chavez,$^{1}$
L. Aguilar,$^{1}$
J. Tomsick,$^{2}$
V. Hambaryan,$^{3}$
and
V. Neustroev$^{4}$}\\
{\small \it %
(1) Institute of Astronomy, UNAM, AP 877, Ensenada, Baja California, 22800 Mexico \\
(2) Space Sciences Laboratory, 7 Gauss Way, UC Berkeley, CA 94720-7450, USA  \\
(3) Astrophys. Institut, Universit\"at Jena, Schillerg\"asschen 2-3, 07745 Jena, Germany   \\
(4)  Centre for Astron., National University of Ireland, Newcastle Rd., Galway,  Ireland   \\
}
}


\AuthorContents{Daisaku Nogami, Paul Mason, and Christian Knigge} 

\AuthorIndex{Nogami}{D.} 

\AuthorIndex{Mason}{P.A.} 

\AuthorIndex{Knigge}{C.} 

     \baselineskip=10pt
     \parindent=10pt

\section*{Abstract} 

We present here results of numerical calculations demonstrating that the very long periodic variability detected in the Cataclysmic Variable FS Aur can be the result of eccentricity modulation of a close binary (CB) orbit induced by the presence of a third body on a circumbinary orbit.  A third component with a substellar mass on a circular, relatively close orbit,  modulates the mass transfer rate of the binary on much longer time scales than periods within the triple system. \\
We also report here preliminary results of X-ray observations of FS Aur, providing further evidence that it contains magnetic and freely precessing white dwarf. These two findings allow us to incorporate the new and previously stressed hypothesis on the nature of FS Aur into one consistent model.

\section{Introduction} 

FS Aur is a short period Dwarf Nova (DN) renown for the presence of a variety of uncommon and largely incomprehensible periodic variabilities of brightness and radial velocity. The orbital period (OP) is determined from the radial velocity (RV) variation of the central peak of emission lines, also referred to as S-wave. The widely used method of making RV measurements using the double-gaussian method reveals a different picture in the case of FS Aur because the wings of the lines are dominated by a periodicity that is longer than the OP, which we call the long spectroscopic period (LSP). The light curve (LC) is even more complicated. The most persistent and dominant LC modulation is the so-called long photometric period (LPP). There is, however, a definite  relation between OP, LSP \& LPP: 1/P$_{ {LSP}} ={{1/P}}_{{OP}} - {{1/P}}_{{LPP}} $,
so the LSP and LPP seem to emerge from the same source within the system. The LC is also highly variable in its appearance, only sometimes showing the OP.  There is also some non-periodic, stochastic variability contaminating the LC on time scales of $1 - 2 \times10^3$\,sec, much longer than the flickering, which is also prominent in the LC.  The amplitudes of the variabilities range from $\le 0.1$ mag flickering to $\approx 0.5$ mag for LPP and stochastic variability.  However, the long term monitoring of FS Aur reveals an additional, discordant, and very long photometric period P$_{\mathrm {VLPP}} \approx 900$ days with an even greater amplitude. In Fig.\ref{fig:lc} up to five $2^m$ dips can be detected in the LC of FS Aur publicly available from  AAVSO (\href{http://www.aavso.org/}), since it became closely monitored during the last decade.  

The optical flux of a DN is dominated by the accretion disk, and the stellar contribution is small or negligible.  Thus, the variability of a DN is mostly the result of internal structure or conditions in the disk or as a result of a variable mass transfer rate. We demonstrate here that the variability with VLPP can be induced by a long-term eccentricity modulation of CB orbit, influencing the mass transfer rate by a third, sub-stellar component orbiting the binary on a close circular orbit with a period much shorter than the observed VLPP.  On the other hand, we will show that the system probably contains a freely precessing magnetic white dwarf (WD) and that LPP and LSP then can form in a coupling region on the inner edge of the accretion disk\cite{Tov1}.

\section{FS Aur as a triple system} 

\begin{figure}[t]

 \begin{tabular}{lr}
  \begin{minipage}{.45\hsize}
   \begin{center}
    \includegraphics[width=7.2cm, bb=0 80 480 320, clip=]{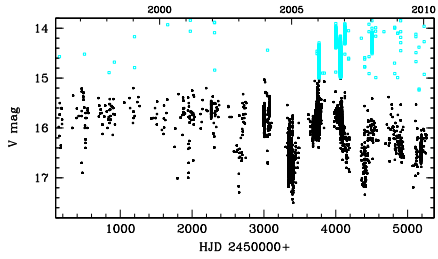}
      \caption{Long-term light curve of FS Aur from the AAVSO data base.  The open (cyan) points correspond to outburst that were duly separated using a V$>15$ mag threshold.  That left some of the outbursts occurring from deep minima unfiltered, but these events are rare and do not affect the period analysis at very low frequencies. The time indicated in the lower axis is in JD and in the upper axis is in years.}
    \label{fig:lc}
   \end{center}
  \end{minipage}

  \begin{minipage}{.49\hsize}
   \begin{center}
    \includegraphics[width=4.8cm, clip=]{tovfig0b.eps}
      \caption{The upper panel shows the VLPP modulation in binary eccentricity as a function of the perturber mass.  The curves correspond to different  $P_{3}/P_{2}$. The thick horizontal line shows the VLPP value. Solutions that cross this line can explain the VLPP. The middle panel shows the perturber mass and orbital radius combinations that result in a long-term modulation of the binary orbit equal to the VLPP. The lower panel shows the amplitude of the binary eccentricity perturbation.}
    \label{fig:Pmodm3}
   \end{center}
  \end{minipage}
 \end{tabular}

\end{figure}

The LC of FS Aur is very complex with large nightly variations sometimes exceeding 0.5 mag. Even with such a large spread, there are also obvious 2 magnitude dips in the average brightness of the system. The dips are sharp and the variability is not sinusoidal, but the succession of minima appears to be periodic (Fig.\ref{fig:lc}). The period determined by the Discrete Fourier Transform (DFT) procedure returns a value around 875 days.  With only five cycles observed during the last 10 years, during which time we had systematic and relatively dense coverage, the period determination is very rough. 
We applied very primitive filters for the observational data, not caring about the precision of the period determination. The irregular outbursts of FS Aur were excluded by deleting points with magnitudes brighter than 15th. Only CCD based observations were taken into account and these were predominantly taken in V-band.  The unfiltered data are included too, since the color difference is negligible compared to the magnitude of the variability. The long nightly observational runs were averaged in the final LC presentation. The period analysis results were not sensitive to whether the nightly data were averaged or not.  In hierarchical triple systems, a third light-weight body can produce perturbations on a central binary, whose period is long compared with the orbital period of the perturber (\cite{Mazeh},\cite{Georgakarakos},\cite{Georgakarakos2}).
 A third companion prevents the complete circularization of the orbit of a binary system by inducing a long-term eccentricity modulation. This long-term modulation is produced by the time-varying tidal force of the perturber upon the binary. We performed numerical analysis to explore the range of possible solutions. The integrations were performed using the high-order Runge-Kutta-Nystr\"om RKN 12(10) 17M integrator of \cite{Brankin} for the full equations of motion in the barycenter frame. 
 At each time step, the instantaneous eccentricity of the binary is computed using the oscillating ellipse approximation (Eq. 2.135 of \cite{Murray}). 
The basic setup is a binary formed by two point masses initially in circular orbit. We used the masses of the binary components common for a 86 min CV (M$_1=0.7M_\odot$ \& M$_2=0.14M_\odot$),  with the orbital period $P_2$ equal to the OP. A third point mass (perturber) moves initially on its own circular orbit, farther away and in the same plane. Its mass $M_3$ and orbital period $P_3$ are changed across an ensemble of numerical experiments. 

The top panel of Figure\,\ref{fig:Pmodm3} shows the resulting periods of the long-term modulation of the binary eccentricity as a function of  the mass of the perturber.  The solid horizontal line corresponds to the VLPP value.  For a perturber whose orbital period is less than 13.4 binary periods, no solution is possible, as their respective curves do not reach up to the VLPP value. For a perturber with periods longer than that but shorter than 23 orbital periods, two solutions are possible: one at low mass and another at an increasingly larger mass.  Finally, a perturber with longer periods produces only one solution at the large mass end of the range.  The curve in the middle panel of Figure\,\ref{fig:Pmodm3} presents the perturber orbit semi-major axis for the solutions that result in the binary system eccentricity modulation of period equal to the observed VLPP value, i.e., solutions which cross the solid line on the upper panel.  The lower panel of Figure\,\ref{fig:Pmodm3} shows the maximum amplitude of the eccentricity perturbation that can be reached for the corresponding solutions presented in the middle panel of this figure. The maximum amplitude is achieved for systems which harbor a third body with M$_3=48$M$_J$.

\begin{figure}[t]
\setlength{\unitlength}{1mm}
\resizebox{10cm}{!}{
\begin{picture}(100,35)(0,0)
\put (0,45){\includegraphics[width=4.6cm, bb=28 26 550 720, angle=-90, clip=]{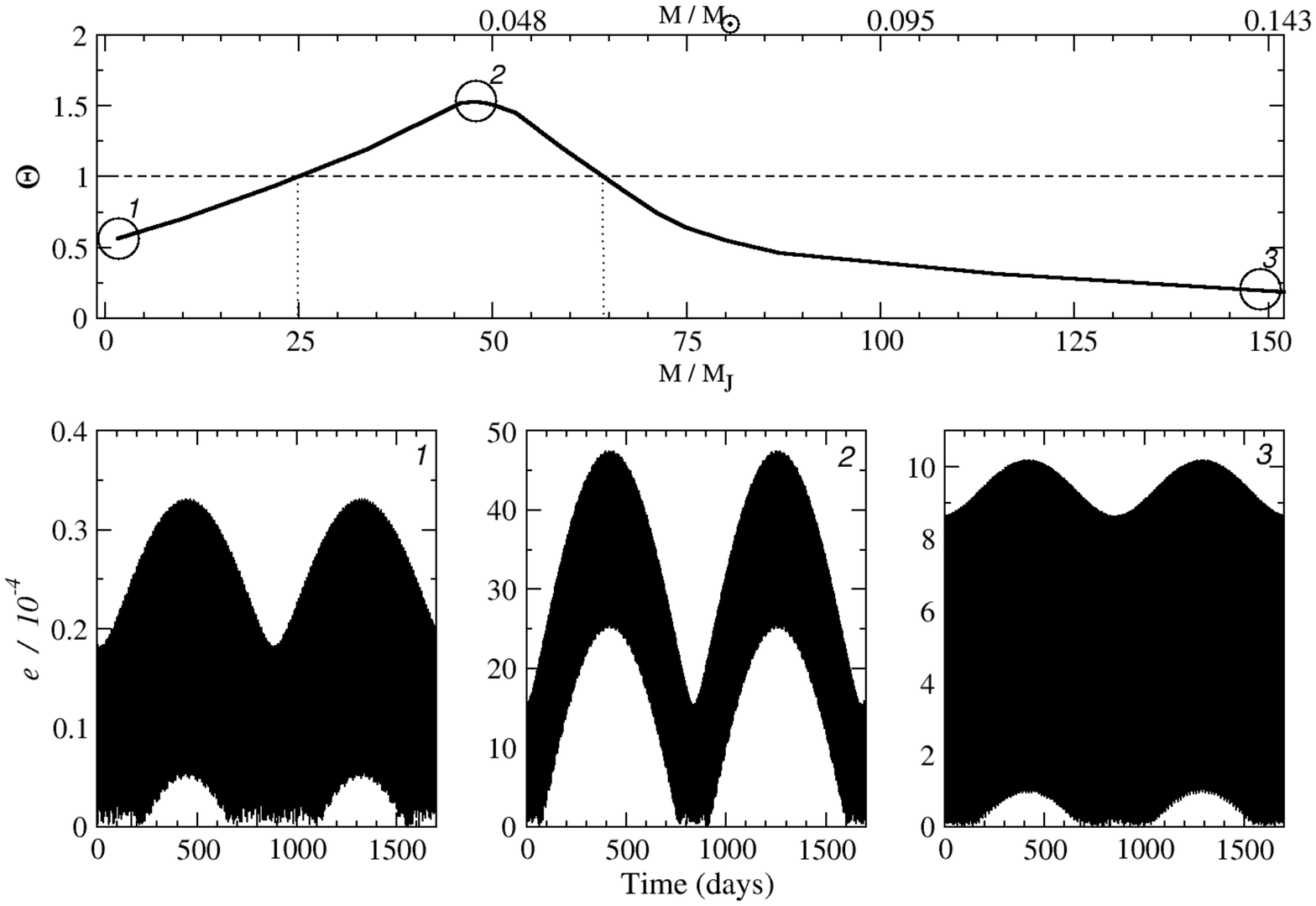}}
\put (62,41){\includegraphics[width=4.15cm, bb = 70 40 550 710, angle=-90, clip=]{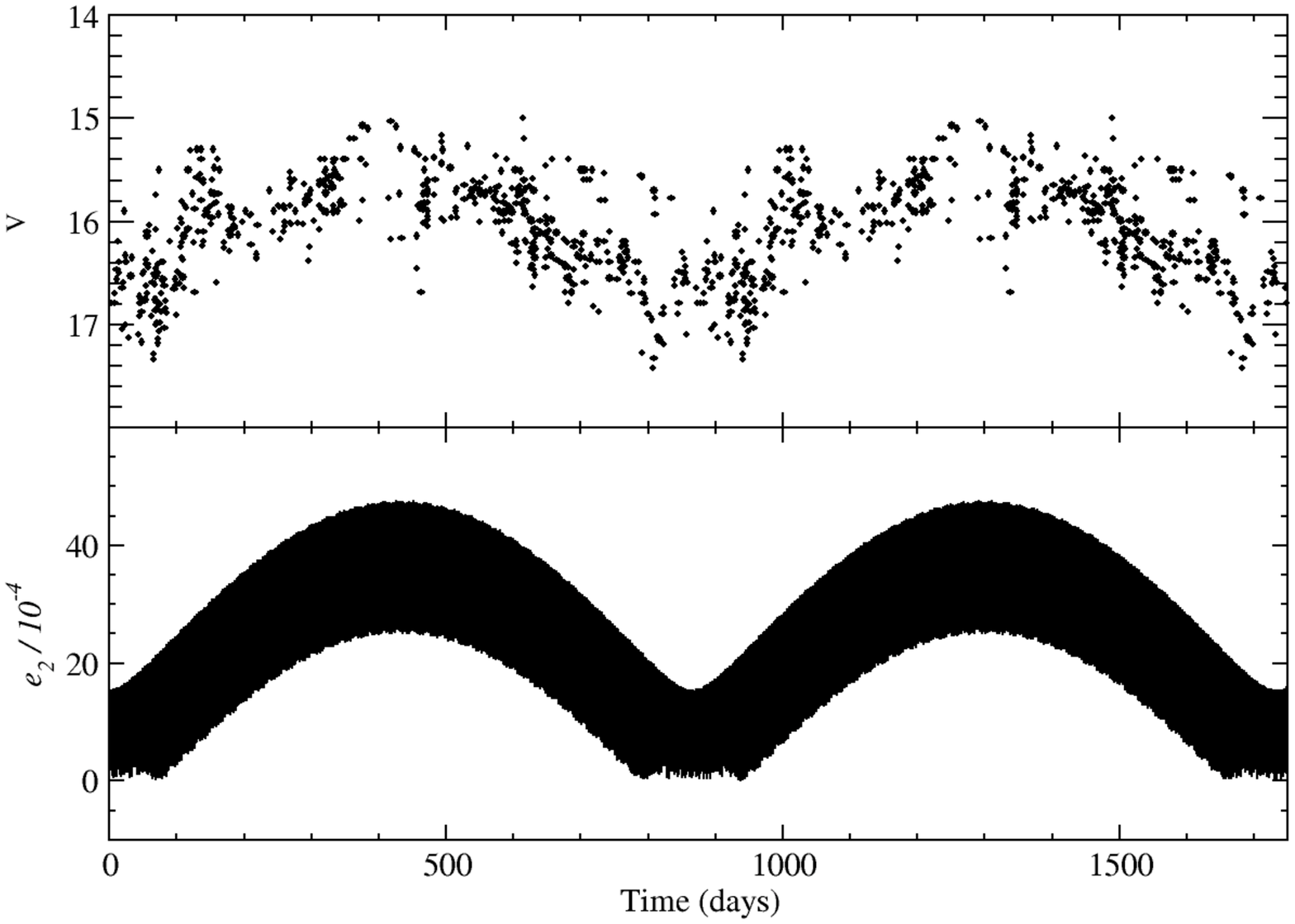}}
\end{picture}}
 \vspace{-0.1pc}
\caption{left)  In the upper panel, the relative amplitude of the low frequency to the high frequency $e_2$ modulations ($\Theta$) is plotted vs. the mass of the perturber.  To show the effect on the shape of the resulting $e_2$ modulation, three extreme cases are presented in the bottom panels: 1) M$_{3}=2$M$_{J}$ and $P_3/P_2=23$; 2) the solution with maximum eccentricity amplitude, with M$_{3}=48$M$_{J}$ and $P_3/P_2=13.4$; 3)  M$_{3}=152$M$_{J}$ and $P_3/P_2=36$. right) The light curve of FS Aur folded with the 875 day period (top panel). In the bottom panel, the eccentricity modulation is shown for the best solution to explain the VLPP (M$_{3}=48$M$_{J}$ and $P_3/P_2=13.4$).}
\label{tovfig1}
\end{figure}

The solutions depend weakly on the exact parameters of the CB, and a wide range of third body masses and orbits will induce the long period effect on the CB eccentricity. More precisely, a third body with $P_3/P_{ {binary}}$ period ratios from 12 to 48, and mass range from a few to a hundred Jupiter masses would be able to produce eccentricity modulation of order of 900 days.  However, the binary eccentricity would be modulated on three different time scales: on the binary period P$_{ {binary}}$, on the period of third body P$_3$ and on the VLPP. The amplitude of each of those modulations would depend primarily on the perturber's mass. The LC folded with 875 day period is presented in the top panel of Fig.\ref{tovfig1}(left). In the bottom panel of Fig.\ref{tovfig1}(right), the modulation of the binary eccentricity is presented for three distinct cases: low, intermediate and high perturber mass (left to right panels).  The amplitude of the low frequency modulation $\Delta e_l$, which basically corresponds to the VLPP, is measured by the maximum to minimum difference of the outline of the corresponding curves. The amplitude of the high frequency modulation $\Delta e_h$, is determined from the width of the strip forming the curves.  The measure of relative amplitude is expressed as a ratio of these amplitudes ($\Theta=\Delta e_{ {l}}/\Delta e_{ {h}}$). The upper panel of Fig.\ref{tovfig1}(right) demonstrates the continuous dependence of the amplitude ratio, $\Theta$, on the mass of the perturber. The three particular cases shown at the bottom of the figure are marked on the curve. It is apparent that the amplitude of the low frequency eccentricity variation is the strongest compared to that of the high frequency, when the perturber's mass hovers around 50 Jupiter masses. Formally, the amplitude of the low frequency modulation becomes dominant ($\Theta \ge 1.0$) in the range of masses for the third body of $25M_J < M_3 < 65M_J$.  The higher mass solutions are also excluded by the absence of any IR excess in the spectral energy distribution of FS Aur. The long term eccentricity modulation curve produced by a third body in such mass range is remarkably similar to the observed LC of FS Aur as can be seen on the lower panel of Fig.\ref{tovfig1}(left). Of course, the physics of translating the eccentricity variation to the brightness of the accretion disc is not that simple, but the eccentricity modulation directly changes the L$_1$ distance from the center of the mass of the secondary and therefore the mass transfer rate. The eccentricity modulations on much shorter time scales have smaller amplitudes, but may be reason for the erratic behavior of FS Aur on comparable time scales. The extended description of numerical calculations and their results and implications upon observed peculiarities of FS Aur are presented in Chavez et al.(2010)\cite{Chavez}.
 
\section{A precessing, magnetic WD in FS Aur }  

Tovmassian et al. (2007) proposed that FS Aur possesses a rapidly rotating magnetic WD precessing with the LPP\cite{Tov1}. Neustroev et al. (2005) intended to reveal the spin period of the WD by fast optical photometry\cite{Ben}. They found a $\sim 100$\,sec peak in the power spectrum, but it was not conclusive evidence after taking into account the weakness of the signal and the peculiarity that it was clearly visible only half of the OP and practically absent in the other half. Nevertheless, Tovmassian et al (2007) discovered the LSP\cite{Tov2} and found no better way to explain it than the same model of the precessing, magnetic WD, which creates a bright source of emission at the inner edge of the truncated accretion disc around the spot where the coupling of the disc matter with the magnetosphere of the WD occurs.  We conducted X-ray observations using Chandra and Swift X-ray telescopes to detect the magnetic WD, determine its spin period and check the proposed model. Here, we report preliminary results of these observations. FS Aur is a relatively bright X-ray source with a count-rate of 0.38 cnts/s as observed with Chandra's ACIS-S3 (back-illuminated) CCD array during a 25~ks exposure. The spectrum is similar to a CV and is best fitted with the multi-temperature mkcflow model with a range of temperatures from 0.3 to 3.8 keV. The spectrum is highly absorbed with the best fit giving a hydrogen column density of $N_H=0.4\times10^{22}\, { {\rm cm}^{-2}}$. The non-magnetic CVs usually have absorption column density of a few times $10^{20}\,{{\rm cm}^{-2}}$\,\cite{Pandel}.  A highly absorbed spectrum is a good indication of a Intermediate Polar (IP) nature for the object\cite{Ramsay}. However, the X-ray LC does not show any clear spin period spike in the power spectrum. There is nevertheless a marginal peak at 101 sec, which is the presumed spin period of WD in FS Aur, and it corresponds to a similarly weak signal in the optical LCs, which might be an indication of the reality of the peak.  The weakness or absence of a spin period signal in X-rays may sound worrisome for claims that FS Aur is an IP and contains a moderately magnetic WD, but Ramsay et al (2008) showed that in the case where the magnetic pole and spin axis are aligned within a few degrees, it will give rise to an X-ray modulation less than a few percent, which would be undetectable due to intrinsic X-ray flickering\cite{Ramsay}.  Actually, the model involving a precessing, magnetic WD in order to explain the LSP would work if the magnetic pole of the WD is close to the spin axis of the star.  

\begin{figure}[t]
\setlength{\unitlength}{1mm}
\resizebox{10cm}{!}{
\begin{picture}(100,42)(0,0)
\put (25,50){\includegraphics[width=4.8cm, bb = 66 30 543 702,  angle=-90, clip=]{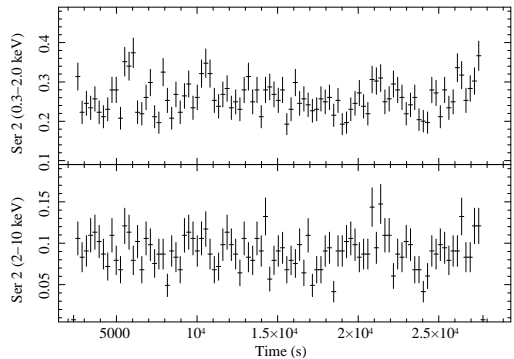}}

\end{picture}}
 \vspace{-1pc}
\caption{The X-ray light curve in soft and hard energy channels.  
}
\label{tovfig2}
\end{figure}

\begin{figure}[t]
\setlength{\unitlength}{1mm}
\resizebox{10cm}{!}{
\begin{picture}(100,34)(0,0)
\put (5,0){\includegraphics[width=5.4cm, bb = 40 75 700 546, clip=]{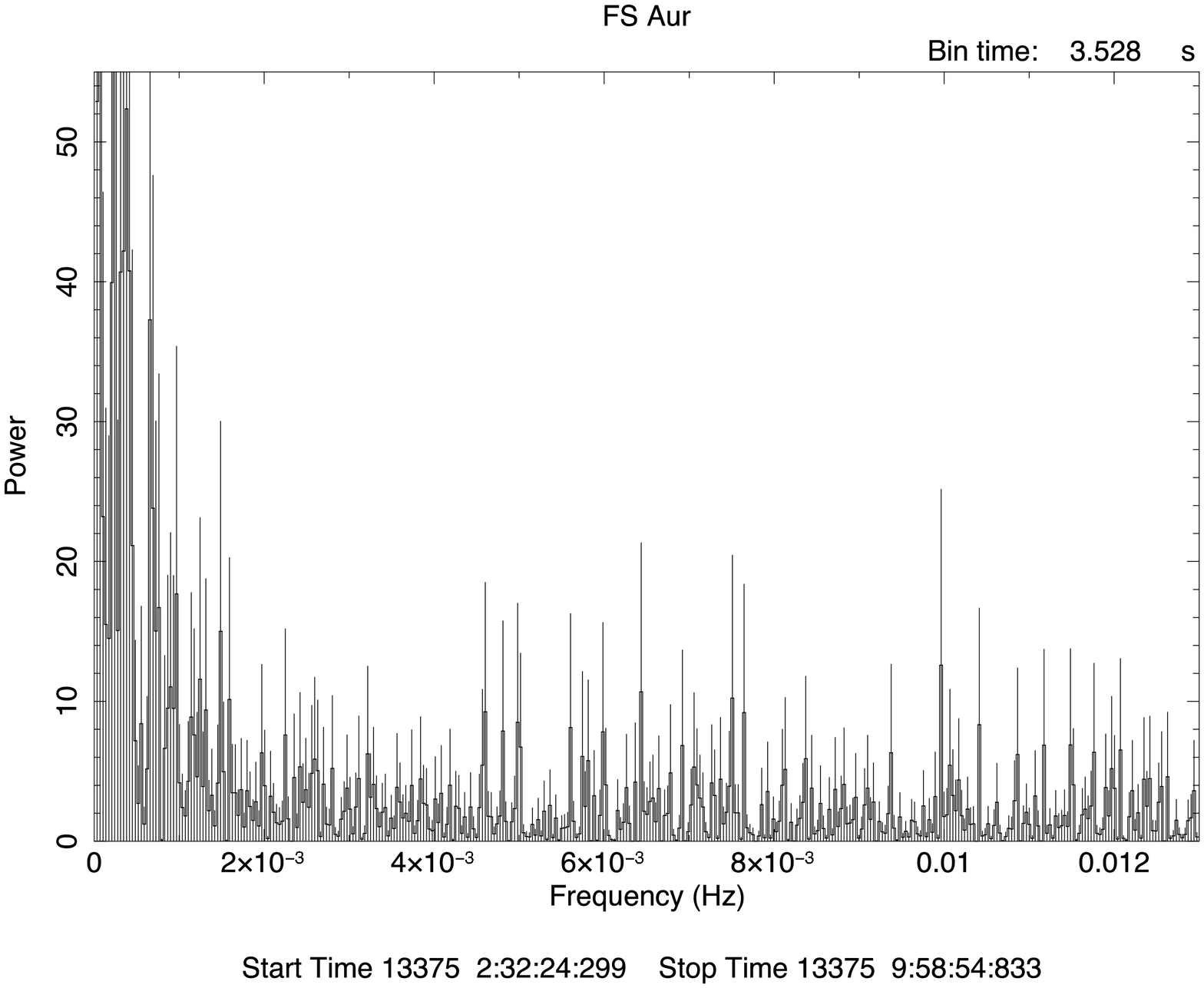}}
\put (62,0){\includegraphics[width=4.0cm, angle=90, clip=]{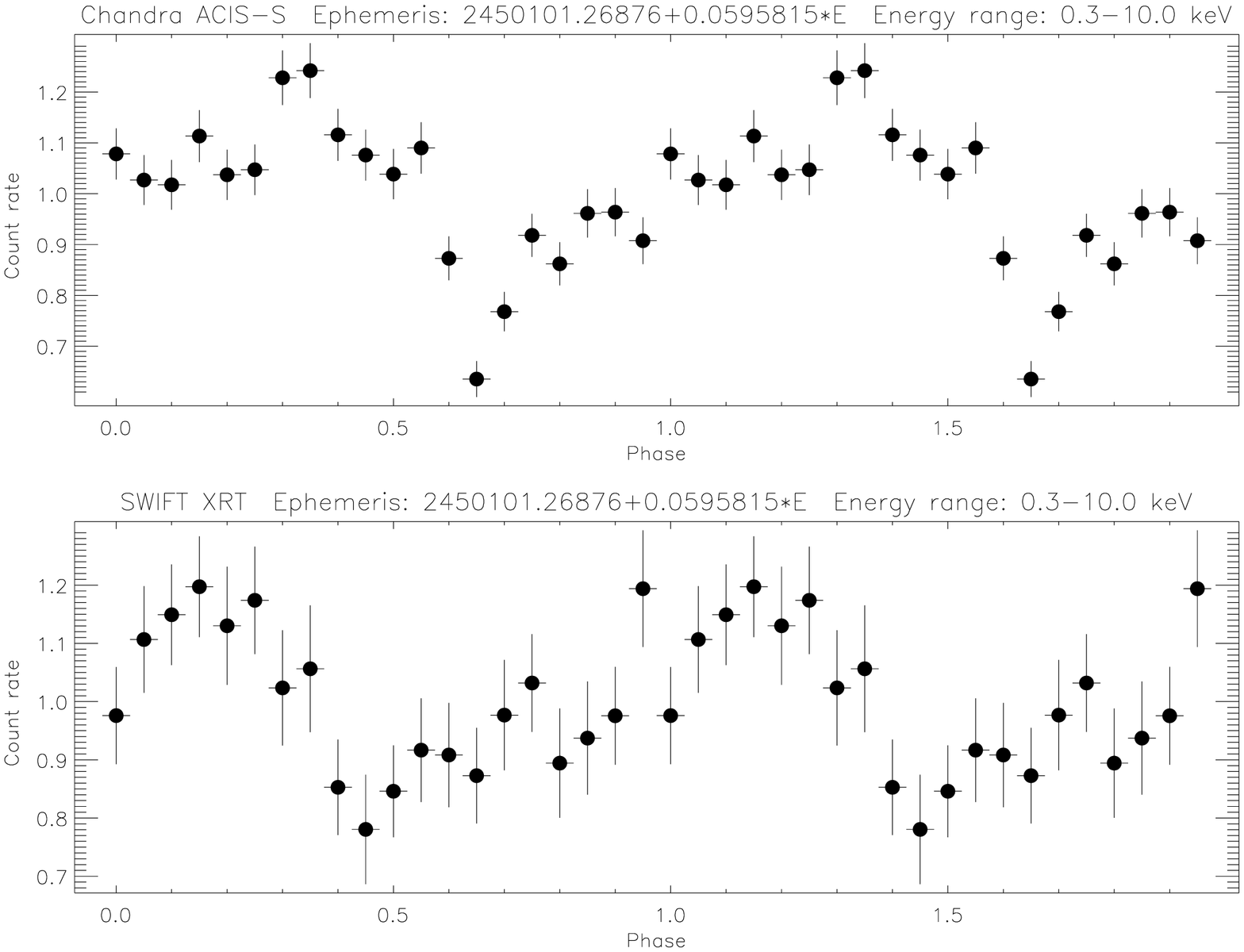}}
\end{picture}}
 \vspace{-0.5pc}
\caption{left) The power spectrum of X-ray light curve.  
right) The light curves from Chandra and Swift folded with the OP and the same T$_0$.}
\label{tovfig3}
\end{figure}

The really important result of the X-ray observations is the detection of OP modulation of the X-ray LC.  As a matter of fact, the X-ray modulation has a pulse profile similar to IPs (i.e., FO Aqr \cite{Beardmore}). OP modulation is detected in almost 50\% of IPs \cite{Parker} with an amplitude of modulation decreasing towards higher energies, and the modulation of FS Aur is stronger in the soft band, indicating photoelectric absorption typical of IPs. In Fig.\ref{tovfig2} the X-ray LC of FS Aur is presented in two energy bands: soft (0.3-2.0 keV) and and hard (2-10 keV).  What is remarkable is that one out of five pulses seems to be missing in the LC.  The Bayesian analysis of photon distribution shows that the absence is statistically significant. In fact, 
 XY Ari exhibits X-ray modulation at the OP that tends to appear and disappear, which they explain with precession, but of tilted/warped disc\cite{Norton}. We believe in case of FS Aur, it is rather evidence supporting the hypothesis of WD precession. Of course, one missing pulse is not good evidence, but follow up X-ray observations with the Swift telescope provided more significant evidence in favor of WD precession. The Swift observations provided shorter continuous segments of the LC, but a larger amount of total coverage. They are comprised of short runs of about $700-1\,200 sec$ each every $\sim 6\,000\ sec$ with total coverage of about 200~ks in three large blocks. The sensitivity of the telescope is lower than that of Chandra (the object is detected as a $0.09\ cnts/sec$ source), but the resulting spectrum and LC are remarkably similar, except the LC has a marked phase shift compared to the LC obtained by Chandra! In Fig.\ref{tovfig3} the X-ray LC of FS Aur folded with the OP is presented. The upper panel presents Chandra data and the lower is Swift data, they are both binned  into equal  time bins with phases calculated from the same zero point determined from the optical spectroscopy. The shift between the two curves is about 0.2 orbital phases. Neither zero point nor the amount of shift have any special relevance; however, the presence of the shift is so far the strongest evidence in favor of precession of the WD in FS Aur. 

\section{Conclusions} 

FS Aur is a 85.797 min DN, which, in addition to the orbital period, demonstrates a number of longer than orbital spectroscopic and photometric periods.  We demonstrate here that in addition to LPP and LSP, the object also shows a very long photometric period, varying up to 2 magnitudes around the quiescent average magnitude with a period of $\sim 900$ days. Such a long period may be explained by the presence of sub-stellar third body on a circular orbit around the CB with orbital period much smaller than the detected VLPP.  The presence of a third body might be helpful in understanding not only the VLPP but also other unusual features in the LC of FS Aur on shorter time scales since the eccentricity modulation occurs on three different time scales. In the case that the mass of third body is somewhere in between 25 and 65\,M$_{\mathrm J}$, the amplitude of the VLPP will be much larger than on short time scales. Among other things, it is probably be more natural to expect a precessing white dwarf in this particular CV, although it is difficult to quantify influence of the third body on a rapidly rotating WD. 

The precession hypothesis, proposed by Tovmassian et al. (2007) finds additional support from X-ray observations of FS Aur. The hard X-ray spectrum and the form of the modulated X-ray LC by the orbital period provide evidence for the suggestion that FS Aur is an IP. In IPs, the orbital modulation is believed to be a result of phase-dependent photoelectric absorption. If we consider that the coupling region of the accretion disk material with the magnetosphere of the WD, where the absorption of X-rays takes place, is the same spot in the inner edge of accretion disk that produces the high RV wings of the emission lines in optical, then the precession hypothesis makes a lot of sense. Since that spot will go around following the precessing magnetic pole of the WD, it will cause phase shifts in the X-ray LC. Thus, we think that the detection of these shifts provides excellent evidence for precession and indirectly confirms that the weak $\sim 100$\,sec peak in power spectra detected in optical and X-rays is indeed the spin period of the WD. 
FS Aur has proved to be very complex, unusual even for a CV, case. It seems that a combination of a triple system containing an IP might resolve most of the issues, even though some mechanisms are not readily understood.


\end{document}